\documentclass[a4paper,12pt]{article}
\usepackage{amsmath,amsthm,amsfonts,amssymb,color}
\usepackage[matrix,arrow]{xy}
\usepackage{graphicx}
\usepackage[UKenglish]{babel}
\usepackage{multicol} 
\usepackage{epstopdf}
\usepackage{hyperref}

\usepackage{caption}
\usepackage{subcaption}

\usepackage{multirow} 
\usepackage{float}
\usepackage{graphicx}
\usepackage{longtable}  

\usepackage{epstopdf}

\usepackage{amsmath,amssymb,amsthm}
\usepackage{enumerate}
\usepackage{amsfonts}
\usepackage{graphics,graphicx}
\usepackage{verbatim}
\usepackage{latexsym,makeidx}

\usepackage{amsmath,amscd}
\usepackage{appendix}
\usepackage[latin1]{inputenc}
\newcommand{\Ste}{{\text{Ste}}}
\newcommand{\Bi}{{\text{Bi}}}
\newcommand{\erf}{{\text{erf}}}

\newenvironment{keywords}{
  \vspace{2mm}
  \noindent
  \keywordsname: 
  \itshape\small
}

\newenvironment{mathsubclass}{
  \small
  \noindent
  \mathsubclassname: 
}

 \def\keywordsname{\textbf{Keywords}}
  \def\mathsubclassname{\textbf{2010 AMS Subject Classification}}

\newtheorem{teo}{Theorem}[section]

\makeatletter
\renewcommand\@biblabel[1]{#1.}
\makeatother
 
\newcommand\be{\begin{equation}}
\newcommand\ee{\end{equation}}

\begin{document}

\title{Integral balance methods applied to a non-classical Stefan problem}

\author{
Julieta Bollati$^{1,2}$,  Mar\'ia Fernanda Natale $^{2}$, Jos\'e A. Semitiel$^{2}$, Domingo A. Tarzia$^{1,2}$\\ \\
\small{$^1$ Consejo Nacional de Investigaciones Cient\'ificas y Tecnol\'ogicas (CONICET)}\\
\small{$^2$ Depto. Matem\'atica - FCE, Univ. Austral, Paraguay 1950} \\  
\small{S2000FZF Rosario, Argentina.} \\
\small{Email: JBollati@austral.edu.ar; FNatale@austal.edu.ar; JSemitiel@austral.edu.ar; DTarzia@austral.edu.ar.}
}

\date{}

\maketitle

\abstract{
In this paper we consider two different  Stefan problems for a semi-infinite material for the non classical heat equation with a source which depends on the heat flux at the fixed face $x=0$. One of them (with constant temperature on $x=0$) was studied in \cite{BrTa} where it was found a unique exact solution of similarity type and the other (with a convective boundary condition at the fixed face) is presented in this work. 
Due to the complexity of the exact solution it is of interest to obtain some kind of approximate solution. For the above reason, the exact solution of each problem is compared with  approximate solutions obtained by applying the heat balance integral method and the refined heat balance  integral method, assuming a quadratic temperature profile in space. In all cases,  a dimensionless analysis is carried out by using the parameters: Stefan number (Ste) and the generalized Biot number (Bi). In addition it is studied the case when Bi goes to infinity, recovering the approximate  solutions  when a Dirichlet condition is imposed at the fixed face. Some numerical simulations are provided in order to verify the accuracy of the approximate methods.
}

\vspace{0.03cm}

\begin{mathsubclass}
35C05, 35C06, 35K05, 35R35, 80A22.
\end{mathsubclass}

\section{Introduction}
Stefan problems model  heat transfer processes that involve a change of phase. They constitute a broad field of study since they arise in  a great number of mathematical and industrial significance problems 
 \cite{AlSo}, \cite{cannon}, \cite{Gu}, \cite{Lu}. A large bibliography on the subject was given in \cite{Ta1} and a review on analytical  solutions  is given in \cite{Ta2}. 
 Nonclassical heat conduction problem for a semi-infinite material was studied in \cite{BTV, CaYi, GlSp, KePr, Ko, TaVi, Vi}.
 A problem of this type involve equations of the form:
\begin{equation}
\dfrac{\partial{U}}{\partial{t}}-\dfrac{\partial ^2U}{\partial x^2}=-F\left(W(t),t\right) ,\qquad  x>0~, \quad  t>0~,
\end{equation}
where $F$ is a given function of two variables. A particular and interesting case is the following
\begin{equation}
F\left(W(t),t\right)=\dfrac{\lambda_{0}}{\sqrt{t}}W(t) \qquad (\lambda_{0}>0)
\end{equation}
where $W=W(t)$ represents the heat flux on the boundary $x=0$, that is $W(t)= \dfrac{\partial{U}}{\partial{x}}(0,t)$.

This kind of problems can be thought of by modelling of a system of temperature regulation in isotropic mediums \cite{TaVi, Vi}, with a nonuniform source term which provides a cooling or heating effect depending upon the properties of $F$ realted to the course of the heat flux at the boundary $x=0$. Other references on the subject are in \cite{GlSp,GlSp2,Ke}.

In this paper, firstly we consider a free boundary problem which consists in determining the temperature $u=u(x,t)$ and the free boundary $x=s(t)$, that we call:

\textbf{Problem $(P)$}. Find the temperature $U=U(x,t)$ at the liquid region $0<x<S(t)$  and the location of the free boundary  $x=S(t)$ such that:
\begin{align}
\rho c \dfrac{\partial{U}}{\partial{t}} - k \dfrac{\partial ^2U}{\partial x^2 }& = - \gamma F\left(\dfrac{\partial{U}}{\partial{x}}(0,t),t \right)~, & 0<x<S(t)~, \qquad  t>0~,\label{Calor}\\
U(0,t)&=u_{\infty}>0~,    &t>0~, \label{t}\\
U(S(t),t)&= 0~,    &t>0~, \label{TempFrontera}\\
k\frac{\partial U}{\partial x}(S(t),t)&=-\rho l \dot{S}(t)~,  &t>0~, \label{CondStefan} \\
S(0)&=0~. &\label{FrontInicial}
\end{align}
where the thermal conductivity  $k$, the mass density  $\rho$, the specific heat $c$  and  the latent heat per unit mass  $l$,  are given constants, $h$ characterizes the heat transfer coefficients \cite{Ta3,ZuCh} and the control function $F$ depend on the the evolution of the heat flux at the boundary $x=0$ as follows:
\begin{equation}
F\left(\dfrac{\partial{U}}{\partial{x}}(0,t),t \right)=\dfrac{\lambda_{0}}{\sqrt{t}}\dfrac{\partial{U}}{\partial{x}}(0,t) \label{FP}
\end{equation}
where $\lambda_{0}>$ is a given constant. This problem was studied in \cite{BrTa}. 

The problem is also considered with another condition at the fixed face $x=0$: the convective condition \cite{tarzia17}. This condition states that heat flux at the fixed face is proportional to the difference between the material temperature and the neighbourhood temperature, that is: \mbox{{ $k\tfrac{\partial U}{\partial x}(0,t)=H(t)\left(U(0,t)-u_{\infty} \right), $}}
where $H(t)$ characterizes the heat transfer at the fixed face and $0<U(0,t)<u_\infty$. In this paper we consider a free boundary problem with a convective condition of the form $H(t)=\frac{h}{\sqrt{t}}, h>0$. 

More precisely, we consider a free boundary problem which consists in determining the temperature $u_h=u_h(x,t)$ and the free boundary $x=s_h(t)$, that we call:

\textbf{Problem ($P_h$)}. Find the temperature $U=U(x,t)$ at the solid region $0<x<S(t)$ and the location of the free boundary $x=S(t)$ such that are satisfied equation (\ref{Calor}), conditions (\ref{TempFrontera})-(\ref{CondStefan})-(\ref{FrontInicial}) of problem (P) and the condition  
\begin{equation}
k\frac{\partial U}{\partial x}(0,t)=\dfrac{h}{\sqrt{t}}(U(0,t)-u_{\infty}),   \qquad t>0~, \label{CondConvecth}
\end{equation}   
instead condition (\ref{t}) of problem $(P)$.

Due to the non-linear nature of this type of problems exact solutions are limited to a few cases  when the exact solution can be find, it is necessary to solve them either numerically or approximately. Despite having the exact solution to the problem that we will study, it is very complicated to find the exact solution. The heat balance integral method introduced by Goodman in \cite{goodman58} is a well-known approximate mathematical technique for solving the location of the free front in heat-conduction problems involving a phase of change. This method consists in transforming the heat equation into an ordinary differential equation over time by assuming a quadratic  temperature profile  in space. In \cite{BoSeTa2018},\cite{Hristov09}, \cite{Hristov09P2}, \cite{Mitchell12}, \cite{MitchellMyers10}, \cite{MitchellMyers10b}, \cite{MitchellMyers12} and in \cite{Mosally02} this method is applied using  different accurate temperature profiles such as: exponential, potential, etc.

Recently, various papers has been published applying integral methods to a variety of thermal and moving boundary problems, especially to  non-linear heat conduction and fractional diffusion: 
\cite{Hr2017, MaMy2016, MiOB2017, Mi2015, Hr2015,FaHr2017, Hr2016, Hr2015-a, Hr2018, Hr2017-a}. 

Different alternative pathways to develop the heat balance integral method were established in \cite{wood01}. In this paper, we obtain approximate solutions through integral heat balance methods and variants obtained thereof proposed in \cite{wood01} for the problems $(P)$ and $(P_{h})$. As one of the mechanisms for the heat conduction is the diffusion, the excitation at the fixed face $x=0$ (for example, a temperature, a flux or a convective condition) does not spread instantaneously to the material \mbox{$x>0$}. However, the effect of the fixed boundary condition can be perceived  in a bounded interval $\left[0,\delta(t)\right]$ (for every time $t>0$) outside of which the temperature remains equal to the initial temperature. The heat balance integral method presented in \cite{goodman58} 
established the existence of a function $\delta=\delta(t)$ that measures the depth of the thermal layer. In problems with a phase of change, this layer is assumed to be the free boundary, i.e $\delta(t)=s(t)$.

From equation (\ref{TempFrontera}) we obtain the new condition:
\begin{equation} \label{StefanAprox}
\left(\dfrac{\partial U}{\partial x}\right)^2(S(t),t)=-\dfrac{l}{kc}\left(\dfrac{\partial ^2 U}{\partial x^2}(S(t),t)-\gamma \frac{\lambda_{0}}{\sqrt{t}}\dfrac{\partial U}{\partial x}(S(t),t)\right)~.
\end{equation}

From equation (\ref{Calor}) and conditions (\ref{TempFrontera})-(\ref{CondStefan}) we obtain the integral condition:
\begin{eqnarray}
\dfrac{d}{dt} \int\limits_{0}^{S(t)} U(x,t) dx&=&-\dfrac{\dfrac{\partial U}{\partial x}(0,t)}{\rho c}\left[\gamma \lambda_{0}\dfrac{S(t)}{\sqrt{t}}+k\right]-\frac{l}{c} \dot{S}(t) ~.\label{EcCalorAproxx}
\end{eqnarray}

The classical heat balance integral method introduced in \cite{goodman58}  to solve problem $(P)$ or $(P_h)$ proposes the resolution of a problem that arises by replacing the equation (\ref{Calor}) by the condition (\ref{EcCalorAproxx}), and the condition (\ref{CondStefan}) by the condition (\ref{StefanAprox}), keeping all others conditions of the problem $(P)$ or $(P_h)$ equals.

In \cite{wood01}, a variant of the classical heat balance integral method was proposed by replacing only equation (\ref{Calor}) by condition (\ref{EcCalorAproxx}), keeping all others conditions of the problem (P) or $(P_h)$ equals.

From equation (\ref{Calor}) and conditions (\ref{t}) and  (\ref{TempFrontera}) we can also obtain:
\begin{eqnarray}
\int\limits_0^{S(t)} \int\limits_0^x \frac{\partial U}{\partial t}(\xi,t) d\xi dx &=& \dfrac{1}{\rho c} \left[-\gamma \lambda_{0}\dfrac{S^{2}(t)}{2\sqrt{t}}\dfrac{\partial U}{\partial x}(0,t)-ku_{\infty}-k\frac{\partial U}{\partial x}(0,t)S(t) \right]~.\label{CalorRIM} 
\end{eqnarray}

The refined  heat balance integral method introduced in \cite{SaSiCo2006} to solve the problem $(P)$ proposes the resolution of the approximate problem that arises by replacing equation (\ref{Calor}) by condition (\ref{CalorRIM}), keeping all others conditions of the problem $(P)$ or $(P_h)$ equals. 

For solving the approximate problems previously defined we propose a quadratic temperature profile in space as follows:
\begin{equation}
U(x,t)=\tilde{A}\left(1-\frac{x}{S(t)}\right)+\tilde{B}\left(1-\frac{x}{S(t)}\right)^2 ,\qquad  0<x<S(t)~, \quad  t>0~,
\end{equation}
where $\tilde{A}$ and $\tilde{B}$ are unknown constants to be determined.

The goal of this paper is to study  different approximations for one-dimensional
one phase Stefan problems with  a  source function that depends on the flux. It is considered two different
problems, which differ from each other in the boundary condition imposed at the fixed face $x =
0$: temperature (Dirichlet) condition or convective (Robin) condition.
In Section 2 we present the exact solution of the problem (P) which was given in \cite{BrTa}. Taking advantage of the exact solution of (P),  we obtain approximate solutions using the heat balance integral method, an alternative method of it and the refined  integral method, comparing each approach with the exact one.
A similar study is done in Section 3 for the problem with a convective condition at the fixed face, (P$_h$). In order to make this analysis, we obtain previously  the exact solution of (P$_h$). 
We also study the limit cases of the obtained approximate solutions when $h\rightarrow \infty$, recovering the approximate solutions when a temperature condition at the fixed face is imposed.

\section{Explicit and approximate solutions to the one-phase Stefan problem for a non-classical heat equation with a source and a temperature condition at the fixed face }
In this section we present the exact solution of the problem (P) and we obtain approximate solutions by using  heat balance integral methods, comparing each approach with the exact one.
\subsection{Exact solution}

In \cite{BrTa},   it  has been proved that for each dimensionless parameter; 
\begin{equation}
\lambda=\frac{\gamma \lambda_0}{\sqrt{k \rho c}}>0 \label{Lambda}
\end{equation}
 the free boundary problem $(P)$,  where $F$  defined by (\ref{FP}), has a unique similarity solution of the type
\begin{align}
u(x,t) & =u_{\infty}\left(1-\dfrac{E(\eta,\lambda)}{E(\xi,\lambda)}\right) \quad,\quad 0< \eta=\dfrac{x}{2a\sqrt{t}}<\xi \\
s(t) & =2a\xi\sqrt{t}\quad,\quad a^{2}=k/\rho c  \;\; \text{(diffusion coefficient)}
\end{align}
where 
\begin{equation}
E(x,\lambda)=\erf(x)+\frac{4 \lambda}{\sqrt{\pi}}\int\limits_{0}^{x} f(r) dr \quad, \quad f(x)=\exp(-x^{2})\int\limits_{0}^{x} \exp(r^{2}) dr  \label{E}
\end{equation}
and $\xi>0$ is the unique solution of
\begin{equation}\label{Xi}
W_{1}(x)=2\lambda W_{2}(x)\quad, \quad x>0
\end{equation}
where the real functions $W_{1}$ and $W_{2}$ are defined by 
\begin{equation}
W_{1}(x)=\Ste \; \exp(-x^{2})-\sqrt{\pi}\;x\; \erf(x) \quad, \quad W_{2}(x)= 2x \int\limits_{0}^{x} f(r) dr -\Ste \;f(x)\;, \label{W1W2}
\end{equation}
$$\erf(x)=\frac{2}{\sqrt{\pi}}\int\limits_0^{x} \exp(-r^2)dr\;,$$
and the dimensionless parameter defined by:
\begin{equation}
\Ste=\frac{cu_{\infty}}{l}
\end{equation}
represent the Stefan number. We remark that function $f$ defined in (\ref{E}), is called the Dawson's integral.

From now on, we will consider the case  $\Ste\in (0,1)$, due to the fact that for  most
phase-change materials candidates over a realistic temperature, the Stefan number will not exceed
1(\cite{So79}).

\subsection{Approximate solution using the classical heat balance integral method}
The classical heat balance integral method in order to solve the problem $(P)$ proposes the resolution of the approximate problem $(P_{1})$ defined as follows:

Find the temperature $u_1=u_1(x,t)$ at the solid region $0<x<s_1(t)$ and the location of the free boundary $x=s_1(t)$ such that satisfy the following conditions: (\ref{t})-(\ref{TempFrontera})-(\ref{FrontInicial})-(\ref{StefanAprox})-(\ref{EcCalorAproxx}).

A solution to this problem will be an approximate one of the problem $(P)$. Proposing the following quadratic temperature
profile in space:
\begin{equation}\label{Temp-u1}
u_1(x,t)=A_1u_\infty\left(1-\frac{x}{s_1(t)}\right)+B_1u_\infty\left(1-\frac{x}{s_1(t)}\right)^2 ,\qquad  0<x<s_1(t)~, \quad  t>0~,
\end{equation}
the free boundary takes the form:
\begin{equation}\label{Front-s1}
s_1(t)=2a\xi_1\sqrt{t},\qquad  t>0~,
\end{equation}
where the constants $A_1$, $B_1$ and $\xi_1$ will be determined from the conditions (\ref{t}), (\ref{StefanAprox}) and (\ref{EcCalorAproxx}). Because of (\ref{Temp-u1}) and (\ref{Front-s1}), the conditions (\ref{TempFrontera}) and (\ref{FrontInicial}) are immediately satisfied. From conditions (\ref{TempFrontera}) and (\ref{EcCalorAproxx}) we obtain: 
\begin{equation}\label{A1}
A_{1}=\dfrac{-2(3+\Ste)\xi_1^2+12\lambda\Ste~ \xi_1+6\Ste}{\Ste\left(\xi_1^2+6\lambda\xi_1+3\right)}~,
\end{equation}
\begin{equation}\label{B1}
B_{1}=\dfrac{3(2+\Ste)\xi_1^2-6\lambda\Ste~ \xi_1-3\Ste}{\Ste\left(\xi_1^2+6\lambda\xi_1+3\right)}~.
\end{equation}

Since $A_1$ and $B_1$ are defined from the parameters $\xi_1$, $\lambda$ and $\Ste$, condition (\ref{StefanAprox}) will be used to find the value of $\xi_1$. In this way, it turns out that $\xi_1$ must be a positive solution of the fifth degree polynomial equation:
\begin{eqnarray}\label{Xi1} 
-4\lambda\left(3+2\Ste\right)z^{5}+2\left(12+9\Ste+2\Ste^2-12\lambda^2(3+2\Ste)\right)z^4-12\lambda\left(-9+16\Ste+4\Ste^4\right)z^3+\nonumber\\
+12(1+2\Ste)\left(-3+(6\lambda^2-1)\Ste\right)z^2+72\lambda\Ste(1+2\Ste)z+18\Ste+3\Ste^2=0~, \quad z>0~. 
\end{eqnarray}

It is easy to see that (\ref{Xi1}) has at least one solution. Descartes' rule of signs states that if the terms of a single-variable polynomial with real coefficients are ordered by descending variable exponent, then the number of positive roots of the polynomial is either equal to the number of sign differences between consecutive nonzero coefficients, or is less than it by an even number. Therefore, in our case, to have a unique root of $(\ref{Xi1})$ is enough to take $\lambda$ such that $12+9\Ste+2\Ste^2-12\lambda^2(3+2\Ste)<0$, that is 
\begin{equation}
\lambda >\sqrt{\frac{2\Ste^{2}+9\Ste+12}{36+24\Ste}}\equiv f(\Ste)
\end{equation}
As $f$ is an increasing function then for $0<\Ste<1$ it is sufficient to take $\lambda > f(1)\cong 0.6191391874$.

Then we have proved the following result:

\begin{teo}\rm The solution of problem (P$_1$), for a quadratic profile in space, is given by (\ref{Temp-u1}) and (\ref{Front-s1}), where the constants $A_1$ and $B_1$ are defined by (\ref{A1}) and (\ref{B1}), respectively and $\xi_1$ is the unique positive solution of the polynomial equation (\ref{Xi1}) if $0<$ Ste $< 1$ and $\lambda >0.62$. 
\end{teo}

As the  approximate methods we are working with are designed as a technique for tracking the location of the free
boundary, the comparisons between the approximate solutions and the exact one will be done on
the free boundary thought the coefficients that characterizes them.  
That is to say, we will compare the known exact solution of the Stefan problem (P) and the approximate solution of the problem (P1) by computing the coefficients $\xi$ and $\xi_1$ that characterizes the free boundaries of each problem, 
which are obtained by solving (\ref{Xi}) and (\ref{Xi1}), respectively. In Figure 1,  we plot the dimensionless coefficients $\xi$ and $\xi_1$ against Stefan number, fixing $\lambda=0.7$. 

\begin{center}
\includegraphics[scale=0.27]{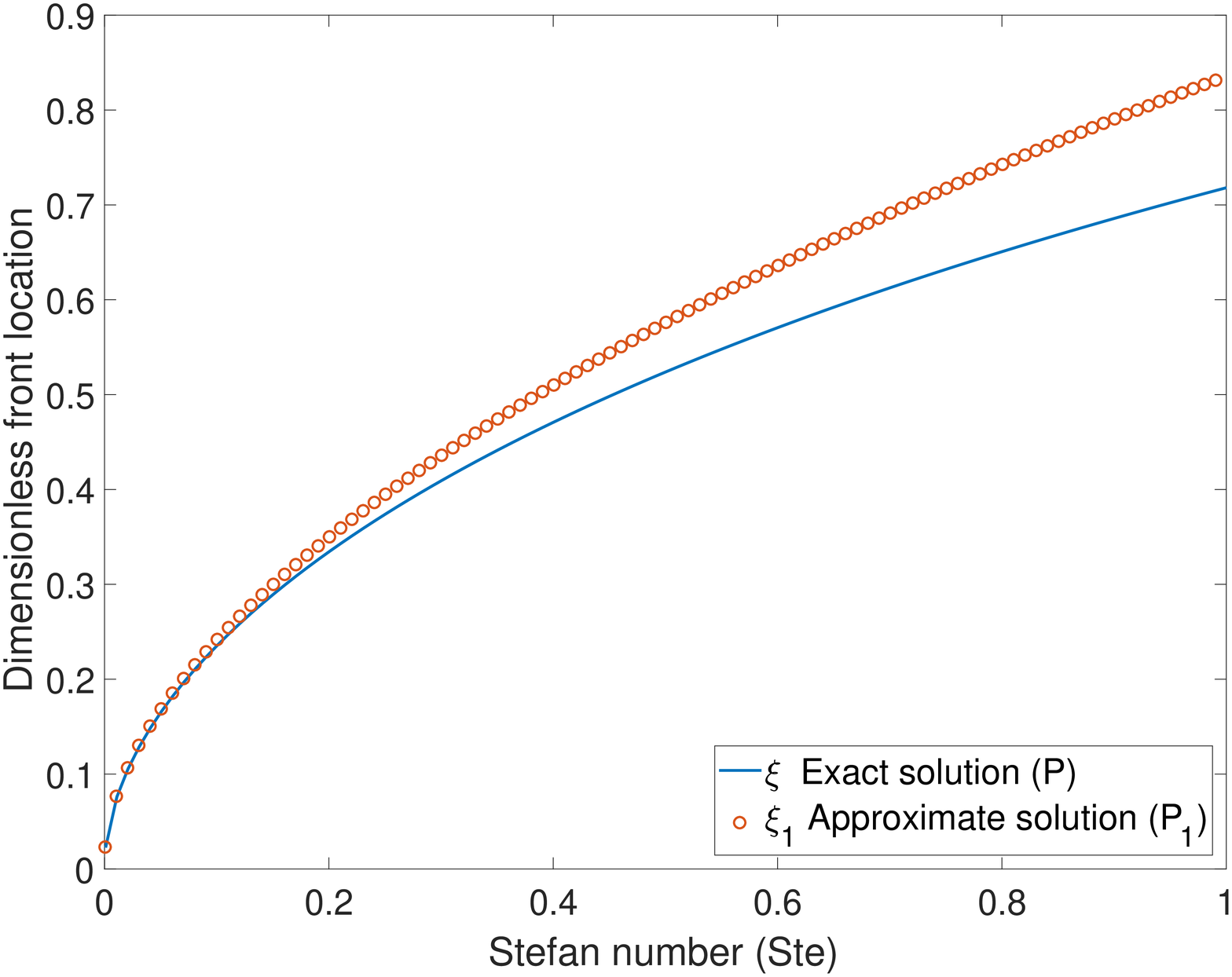} \\

{\scriptsize Figure 1: Plot of the dimensionless coefficients $\xi$ and $\xi_1$  against Ste number, for $\lambda=0.7$.}

\end{center}

\subsection{Approximate solution using a modified method of the classical heat balance method}
An alternative method of the classical heat balance integral method in order to solve the problem $(P)$ proposes the resolution of the approximate problem (P$_2$) defined as follows:

Find the temperature $u_2=u_2(x,t)$ at the solid region $0<x<s_2(t)$ and the location of the free boundary $x=s_2(t)$ such that satisfy the following conditions: (\ref{t})-(\ref{TempFrontera})-(\ref{CondStefan})-(\ref{FrontInicial})-(\ref{EcCalorAproxx}).

A solution to this problem will be an approximate one of the problem (P). Proposing the following quadratic temperature
profile in space:
\begin{equation}\label{Temp-u2}
u_2(x,t)=A_2u_\infty\left(1-\frac{x}{s_2(t)}\right)+B_2u_\infty\left(1-\frac{x}{s_2(t)}\right)^2 ,\qquad  0<x<s_2(t)~, \quad  t>0~,
\end{equation}
the free boundary takes the form:
\begin{equation}\label{Front-s2}
s_2(t)=2a\xi_2\sqrt{t},\qquad  t>0~,
\end{equation}
where the constants $A_2$, $B_2$ and $\xi_2$ will be determined from the conditions (\ref{t}), (\ref{CondStefan}) and (\ref{EcCalorAproxx}). Because of (\ref{Temp-u2}) and (\ref{Front-s2}), the conditions (\ref{TempFrontera}) and (\ref{FrontInicial}) are immediately satisfied. From conditions (\ref{TempFrontera}) and (\ref{EcCalorAproxx}) we obtain: 
\begin{equation}\label{A2}
A_{2}=\dfrac{2}{\Ste}\xi_2^2~,
\end{equation}
\begin{equation}\label{B2}
B_{2}=1-\dfrac{2}{\Ste}\xi_2^2~.
\end{equation}

Since $A_2$ and $B_2$ are defined from the parameters $\xi_2$ and $\Ste$, condition (\ref{CondStefan}) will be used to find the value of $\xi_2$. In this way, it turns out that $\xi_2$ must be a positive solution of the fourth degree polynomial equation:
\begin{eqnarray}\label{Xi2} 
z^4+6\lambda z^3+\left(6+\Ste\right)z^2-6\lambda\Ste~z-3\Ste=0~, \qquad z>0~. 
\end{eqnarray}
It is easy to see, using the Descartes' rule, that (\ref{Xi2}) has a unique positive  solution.

Therefore, the following theorem holds:

\begin{teo}\rm The solution of problem (P$_2$), for a quadratic profile in space, is given by (\ref{Temp-u2}) and (\ref{Front-s2}), where the constants $A_2$ and $B_2$ are defined by (\ref{A2}) and (\ref{B2}), respectively and $\xi_2$ is the unique positive solution of the polynomial equation (\ref{Xi2}). 
\end{teo}

To compare the free boundaries obtained in problem (P) and the approximate problem (P$_2$), we compute the coefficient that characterizes the free boundaries. The exact value of $\xi$ and the approach $\xi_2$ are the unique roots of equations (\ref{Xi}) and (\ref{Xi2}), respectively.

 Figure 2 shows, for Stefan values up to 1, how the dimensionless coefficient $\xi_2$, which characterizes
the location of the free boundary $s_2$, approaches the coefficient $\xi$, corresponding to the exact free
boundary $s$, when the dimensionless parameter is $\lambda=0.7$.

\begin{center}
\includegraphics[scale=0.27]{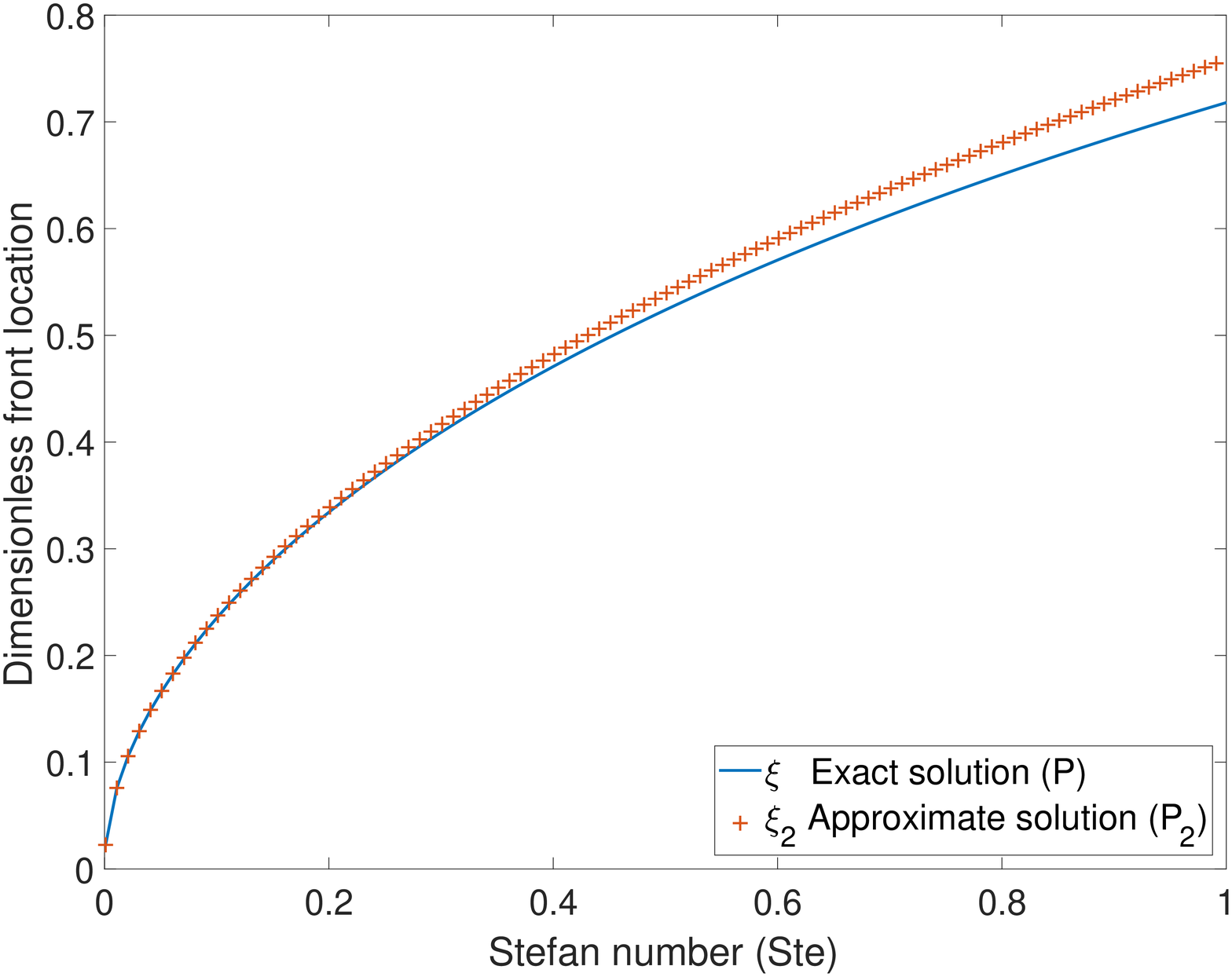} \\
{\scriptsize Figure 2: Plot of the dimensionless coefficients $\xi$ and $\xi_2$  against Ste number, for $\lambda=0.7$.}

\end{center}

\subsection{Approximate solution using the refined integral method}
The refined heat balance integral method in order to solve the problem (P), proposes the resolution of an approximate problem (P$_3$) formulated as follows:

Find the temperature $u_3=u_3(x,t)$ at the solid region $0<x<s_3(t)$ and the location of the free boundary $x=s_3(t)$ such that satisfy the following conditions: (\ref{t})-(\ref{TempFrontera})-(\ref{CondStefan})-(\ref{FrontInicial})-(\ref{CalorRIM}).

A solution to this problem will be an approximate one of the problem (P). Proposing the following quadratic temperature
profile in space:
\begin{equation}\label{Temp-u3}
u_3(x,t)=A_3u_\infty\left(1-\frac{x}{s_3(t)}\right)+B_3u_\infty\left(1-\frac{x}{s_3(t)}\right)^2 ,\qquad  0<x<s_3(t)~, \quad  t>0~,
\end{equation}
the free boundary takes the form:
\begin{equation}\label{Front-s3}
s_3(t)=2a\xi_3\sqrt{t},\qquad  t>0~,
\end{equation}
where the constants $A_3$, $B_3$ and $\xi_3$ will be determined from the conditions (\ref{t}), (\ref{CondStefan}) and (\ref{CalorRIM}). Because of (\ref{Temp-u3}) and (\ref{Front-s3}), the conditions (\ref{TempFrontera}) and (\ref{FrontInicial}) are immediately satisfied. From conditions (\ref{TempFrontera}) and (\ref{CalorRIM}) we obtain: 
\begin{equation}\label{A3}
A_{3}=\dfrac{2}{\Ste}\xi_3^2~,
\end{equation}
\begin{equation}\label{B3}
B_{3}=1-\dfrac{2}{\Ste}\xi_3^2~.
\end{equation}

Since $A_3$ and $B_3$ are defined from the parameters $\xi_3$ and $\Ste$, condition (\ref{CondStefan}) will be used to find the value of $\xi_3$. In this way, it turns out that $\xi_3$ must be a positive solution of the third degree polynomial equation:
\begin{eqnarray}\label{Xi3} 
-6\lambda z^3-\left(6+\Ste\right)z^2+6\lambda \Ste~z+3\Ste=0~, \qquad z>0~. 
\end{eqnarray}
It is easy to see, using the Descartes' rule, that (\ref{Xi3}) has a unique positive solution.

Therefore, the following theorem holds:
\begin{teo}\rm The solution of problem (P$_3$), for a quadratic profile in space, is given by (\ref{Temp-u3}) and (\ref{Front-s3}), where the constants $A_3$ and $B_3$ are defined by (\ref{A3}) and (\ref{B3}), respectively and $\xi_3$ is the unique positive solution of the polynomial equation (\ref{Xi3}). 
\end{teo}

To compare the free boundaries obtained in problem (P) and the approximate problem (P$_3$), we compute the coefficient that characterizes the free boundaries. The exact value of $\xi$ and the approach $\xi_3$ is obtained by solving the equations obtained in (\ref{Xi}) and (\ref{Xi3}), respectively.

For every $\Ste < 1$, we plot the numerical value of the dimensionless coefficient $\xi_3$ obtained by applying the refined integral method, 
against the exact coefficient $\xi$ (Figure 3).

\begin{center}
\includegraphics[scale=0.27]{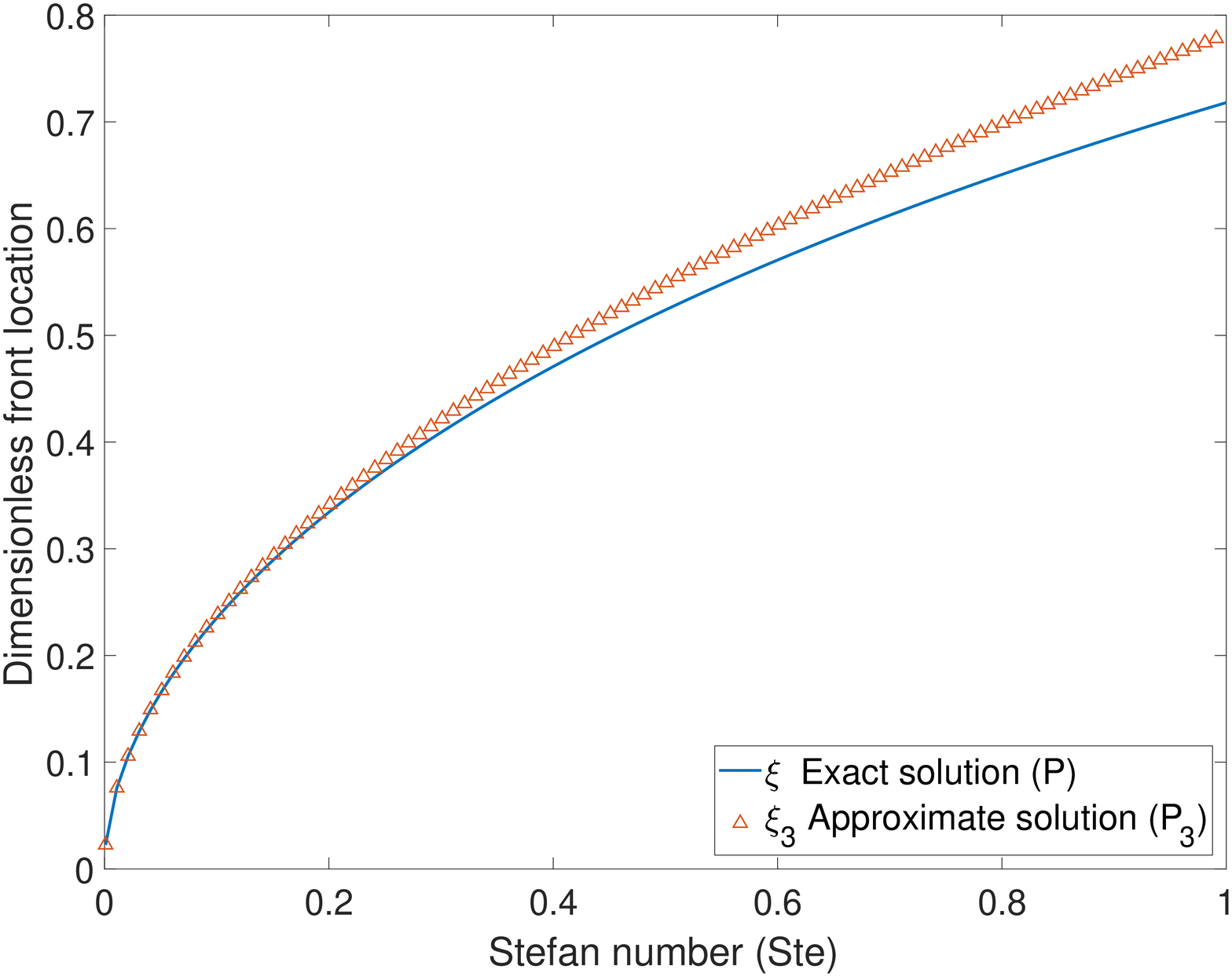} \\
{\scriptsize Figure 3: Plot of the dimensionless coefficients $\xi$ and $\xi_3$  against Ste number, for $\lambda=0.7$.}

\end{center}

\subsection{Comparisons between solutions}

In this subsection, for different Ste numbers, we make comparisons between the numerical value of the coefficient $\xi$ given by (\ref{Xi})  and the approximations $\xi_1$, $\xi_2$ and $\xi_3$ given by (\ref{Xi1}), (\ref{Xi2}), (\ref{Xi3}), respectively.  In order to compare the approximate solution with the exact one, and to obtain which technique gives the best agreement, we display in Table 1, varying Ste between 0 and 1, the exact dimensionless free front, and its different approaches, showing also the percentage relative error committed in each case being $E_{\text{rel}}(\xi_i)=100 \frac{|\xi_i-\xi|}{|\xi|}$.

\begin{center}

\begin{footnotesize}

Table 1: Dimensionless free front coefficients and its relative errors for $\lambda=0.7$.\\[0.15cm]
\begin{tabular}{cc|cc|cc|cc}
\hline
Ste      & $\xi$ (P)     &  $\xi_1$  (P$_1$) & $E_{\text{rel}}(\xi_1)$       &   $\xi_2$  (P$_2$)   & $E_{\text{rel}}(\xi_2)$      &    $\xi_3$  (P$_3$) & $E_{\text{rel}}(\xi_3)$   \\
\hline

    0.1 &    0.2351  &  0.2401  \quad &  2.139 \%             &   0.2363  &  0.545  \%  &   0.2373  &  0.963 \% \\
    0.2 &    0.3342  &  0.3486  \quad &  4.289 \%             &   0.3381  &  1.145  \%  &   0.3408  &  1.956 \% \\
    0.3 &    0.4091  &  0.4348  \quad &  6.284 \%             &   0.4162  &  1.753  \%  &   0.4211  &  2.932 \% \\
    0.4 &    0.4708  &  0.5089  \quad &  8.109 \%             &   0.4818  &  2.351  \%  &   0.4890  &  3.877 \% \\
    0.5 &    0.5238  &  0.5750  \quad &  9.776 \%             &   0.5392  &  2.934  \%  &   0.5489  &  4.788 \% \\
    0.6 &    0.5706  &  0.6350  \quad &  11.30 \%             &   0.5905  &  3.501  \%  &   0.6029  &  5.666 \% \\
    0.7 &    0.6125  &  0.6903  \quad &  12.69 \%             &   0.6373  &  4.048  \%  &   0.6524  &  6.510 \% \\
    0.8 &    0.6507  &  0.7417  \quad &  13.98 \%             &   0.6805  &  4.577  \%  &   0.6983  &  7.322 \% \\
    0.9 &    0.6857  &  0.7897  \quad &  15.16 \%             &   0.7206  &  5.087  \%  &   0.7413  &  8.102 \% \\
    1.0 &    0.7181  &  0.8348  \quad &  16.25 \%             &   0.7582  &  5.579  \%  &   0.7817  &  8.854 \% \\
    \hline
\end{tabular}

\end{footnotesize}

\end{center}

 It may be noticed that the relative error committed in each approximate technique increases when the Stefan number becomes greater, reaching the percentages   16.25 \% , 5.579  \% and  8.854 \%  for the problems  (P$_1$), (P$_2$) and (P$_3$) respectively.

\section{Explicit and approximate solutions to the one-phase Stefan problem for a non-classical heat equation with a source and a convective condition at the fixed face}

In this section we present the exact solution of the problem (P$_h$) and we obtain different approaches by using  heat balance integral methods, comparing them with the exact one.

\subsection{Exact solution}

In this subsection we will obtain the exact solution of the problem (P$_h$) given by (\ref{Calor}),(\ref{TempFrontera})-(\ref{FrontInicial}) and (\ref{CondConvecth})  instead of condition (\ref{t}) of problem $(P)$.

In similar way as \cite{BrTa}, if  we define the similarity variable $\eta=\tfrac{x}{2a\sqrt{t}}$ and $\Phi(\eta)=u_h(x,t)$, then (P$_h$) turns equivalent to the following ordinary differential problem:
\begin{align}
\Phi''(\eta)+2\eta \Phi'(\eta)&= 2\lambda \Phi'(0),\qquad 0<\eta<\xi_h \label{Phi-1} \\
\Phi'(0)&=2 \text{Bi} \left(\Phi(0)-u_{\infty} \right) \\
\Phi(\xi_h)&=0\\
\Phi'(\xi_h)&=-2\frac{u_{\infty}}{\Ste} \xi_h \label{Phi-4}
\end{align}
where the dimensionless parameter defined by 
\begin{equation}
\Bi=\dfrac{h a}{k} \label{Bi}
\end{equation}
represent the  Biot number and $\xi_h$ is the coefficient that characterizes the free boundary $s_h$.

It is a simple matter to find the solution to (\ref{Phi-1})-(\ref{Phi-4}) and thus the solution to (P$_h$) which is given by
\begin{align}
u_h(x,t)&=\Phi(\eta)=\dfrac{\Bi u_{\infty} \sqrt{\pi}}{1+\Bi \sqrt{\pi} E(\xi_h,\lambda)} \left[ E(\xi_h,\lambda)-E(\eta,\lambda)\right], \qquad 0<\eta<\xi_h \\
s_h(t) & =2a\xi_h\sqrt{t}
\end{align}
where the function $E$ is given by (\ref{E}) and $\xi_h$ must be a solution of 
\begin{equation}
W_{1h}(x)=2\lambda W_2(x), \qquad x>0 \label{Xih}
\end{equation}
with $W_{1h}(x)=W_1(x)-\dfrac{kx}{h a}$,   and $W_1$, $W_2$ given by (\ref{W1W2}).

It can be proved that $W_{1h}$ has the same properties as $W_1$, therefore we can apply the results obtained in \cite{BrTa} to conclude that there exists a unique solution $\xi_h$ of (\ref{Xih}).

Notice that in problem (P$_h$) a convective boundary condition (\ref{CondConvecth}) characterized by the coefficient $h$ at the fixed face $x=0$ is imposed. This condition constitutes a generalization of the Dirichlet one in the sense that if we take de limit when $h \rightarrow \infty$ we must obtain $U(0,t)=u_\infty$. From definition (\ref{Bi}), studying the limit behaviour of the solution to our problem (P$_h$) when $h \rightarrow \infty$ is equivalent to study the case when $\Bi \rightarrow \infty$.

If for every $h$, we define $\xi_h$ as the unique solution to (\ref{Xih}) then, it can be observed that $\lbrace \xi_h \rbrace$ is increasing and bounded, and so convergent. In addition, it can be easily seen that $\xi_h\to\xi$ where $\xi$ is the unique solution to (\ref{Xi}). Then, we can state the following result:

\begin{teo}\rm The solution to problem (P$_h$) converges to the solution to problem (P) when \mbox{$Bi \rightarrow \infty$} \; (i.e. $h\to\infty$), that is:
\begin{eqnarray*}
\lim\limits_{h\to \infty} s_h(t) &=& s(t), \qquad  t>0 \\
\lim\limits_{h\to \infty} u_h(x,t) &=& u(x,t), \qquad  0<x<s(t), \; t>0  \\
\end{eqnarray*}
\end{teo}

\subsection{Approximate solution using the classical heat balance integral method}
The classical heat balance integral method in order to solve the problem (P$_h$) proposes the resolution of the approximate problem  (P$_{h_1}$) defined as follows:

Find the temperature $u_{h_1}=u_{h_1}(x,t)$ at the solid region $0<x<s_{h_1}(t)$ and the location of the free boundary $x=s_{h_1}(t)$ such that satisfy the following conditions: (\ref{TempFrontera})-(\ref{FrontInicial})-(\ref{CondConvecth})-(\ref{StefanAprox})-(\ref{EcCalorAproxx}).

A solution to this problem will be an approximate one of the problem (P$_h$). Proposing the following quadratic temperature
profile in space:
\begin{equation}\label{Temp-uh1}
u_{h_1}(x,t)=A_{h_1}u_\infty\left(1-\frac{x}{s_{h_1}(t)}\right)+B_{h_1}u_\infty\left(1-\frac{x}{s_{h_1}(t)}\right)^2 ,\qquad  0<x<s_{h_1}(t)~, \quad  t>0~,
\end{equation}
the free boundary takes the form:
\begin{equation}\label{Front-sh1}
s_{h_1}(t)=2a\xi_{h_1}\sqrt{t},\qquad  t>0~,
\end{equation}
where the constants $A_{h_1}$, $B_{h_1}$ and $\xi_{h_1}$ will be determined from the conditions (\ref{CondConvecth}), (\ref{StefanAprox}) and (\ref{EcCalorAproxx}). Because of (\ref{Temp-uh1}) and (\ref{Front-sh1}), the conditions (\ref{TempFrontera}) and (\ref{FrontInicial}) are immediately satisfied. From conditions (\ref{TempFrontera}) and (\ref{EcCalorAproxx}) we obtain: 
\begin{equation}\label{Ah1}
A_{h_1}=\dfrac{-2(3+\Ste)\xi_{h_1}^2+\left(12\lambda\Ste-\frac{6}{\Bi}\right) \xi_{h_1}+6\Ste}{\Ste\left(\xi_{h_1}^2+\left(6\lambda+\frac{2}{\Bi}\right)\xi_{h_1}+3\right)}~,
\end{equation}
\begin{equation}\label{Bh1}
B_{h_1}=\dfrac{3(2+\Ste)\xi_{h_1}^2+\left(\frac{3}{\Bi}-6\lambda\Ste\right) \xi_{h_1}-3\Ste}{\Ste\left(\xi_{h_1}^2+\left(6\lambda+\frac{2}{\Bi}\right)\xi_{h_1}+3\right)}~.
\end{equation}

Since $A_{h_1}$ and $B_{h_1}$ are defined from the parameters $\xi_{h_1}$, $\lambda$, $\Ste$ and $\Bi$, condition (\ref{StefanAprox}) will be used to find the value of $\xi_{h_1}$. In this way, it turns out that $\xi_{h_1}$ must be a positive solution of the fifth degree polynomial equation:
\begin{eqnarray}\label{Xih1} 
-4\lambda\left(3+2\Ste\right)z^{5}+2\left(12+9\Ste+2\Ste^2-12\lambda^2(3+2\Ste)-\frac{4\lambda}{\Bi}(3+2\Ste)\right)z^4+\nonumber\\
+\left(-12\lambda(-9+16\Ste+4\Ste^4)+\frac{6}{\Bi}(7+2\Ste)\right)z^3+\nonumber\\
+12\left((1+2\Ste)(-3+(6\lambda^2-1)\Ste)+\frac{2}{\Bi^2}-\frac{3\lambda}{\Bi}(1+\Ste)\right)z^2+ \nonumber\\
+\left(72\lambda\Ste(1+2\Ste)-\frac{6}{\Bi}(3+10\Ste)\right)z+18\Ste+3\Ste^2=0~, \qquad z>0~. 
\end{eqnarray}


It is easy to see that equation (\ref{Xih1}) has at least one solution. In order to prove uniqueness, we are going to use Descartes rule. Therefore, if we rewrite (\ref{Xih1}) as $\sum\limits_{i=0}^{5} a_i z^i=0$, we have to analyse the sign of each coefficient $a_i$.

Clearly, $a_5<0$ and $a_0>0$.  For $0<\Ste<1$ and $\lambda>0.62$, as in problem (P$_1$), $a_4<0$ for all $\Bi$. Under these hypothesis:  $a_3<0$ if and only if $\Bi>\frac{7+2\Ste}{2\left(9\lambda +16 \lambda \Ste +4\lambda\Ste^2\right)}$, and  $a_1>0$ if and only if $\Bi>\frac{3+10\Ste}{12\lambda\Ste\left(1+2\Ste\right)}$. Summarizing what has been discussed, the following theorem holds:

\begin{teo}\rm The solution of problem (P$_{h_1}$), for a quadratic profile in space, is given by (\ref{Temp-uh1}) and (\ref{Front-sh1}), where the constants $A_{h_1}$ and $B_{h_1}$ are defined by (\ref{Ah1}) and (\ref{Bh1}), respectively and $\xi_{h_1}$ is the unique positive solution of the polynomial equation (\ref{Xih1}) if
 $$0<\Ste<1, \qquad \lambda>0.62 \quad  \text{and }\quad \Bi>\frac{3+10\Ste}{12 \lambda \Ste (1+2 \Ste)}.$$
In addition, the solution to problem (P$_{h_1}$) converges to the solution to problem (P$_1$) when $\Bi \rightarrow \infty$.
\end{teo}

To compare the solutions obtained in (P$_h$) and (P$_{h_1}$), we compute the coefficient that characterizes the free boundary in each problem. The exact value of $\xi_{h}$ and the approach $\xi_{h_1}$ are obtained by solving the equations obtained in (\ref{Xih}) and (\ref{Xih1}), respectively.  In Figure 4 we plot  the coefficients $\xi_h$ and $\xi_{h_1}$ against Bi in order to visualize the behaviour of the approximate solution, fixing $\Ste=0.5$ and $\lambda=0.7$. In order that the convergence mentioned above of $\xi_h\to  \xi$ and $\xi_{h_1}\to \xi_1$ when $\Bi\to\infty$, could be appreciated, we also plot $\xi$ and $\xi_1$ given by the solution of (\ref{Xi}) and (\ref{Xi1}), respectively.

\begin{center}
\includegraphics[scale=0.27]{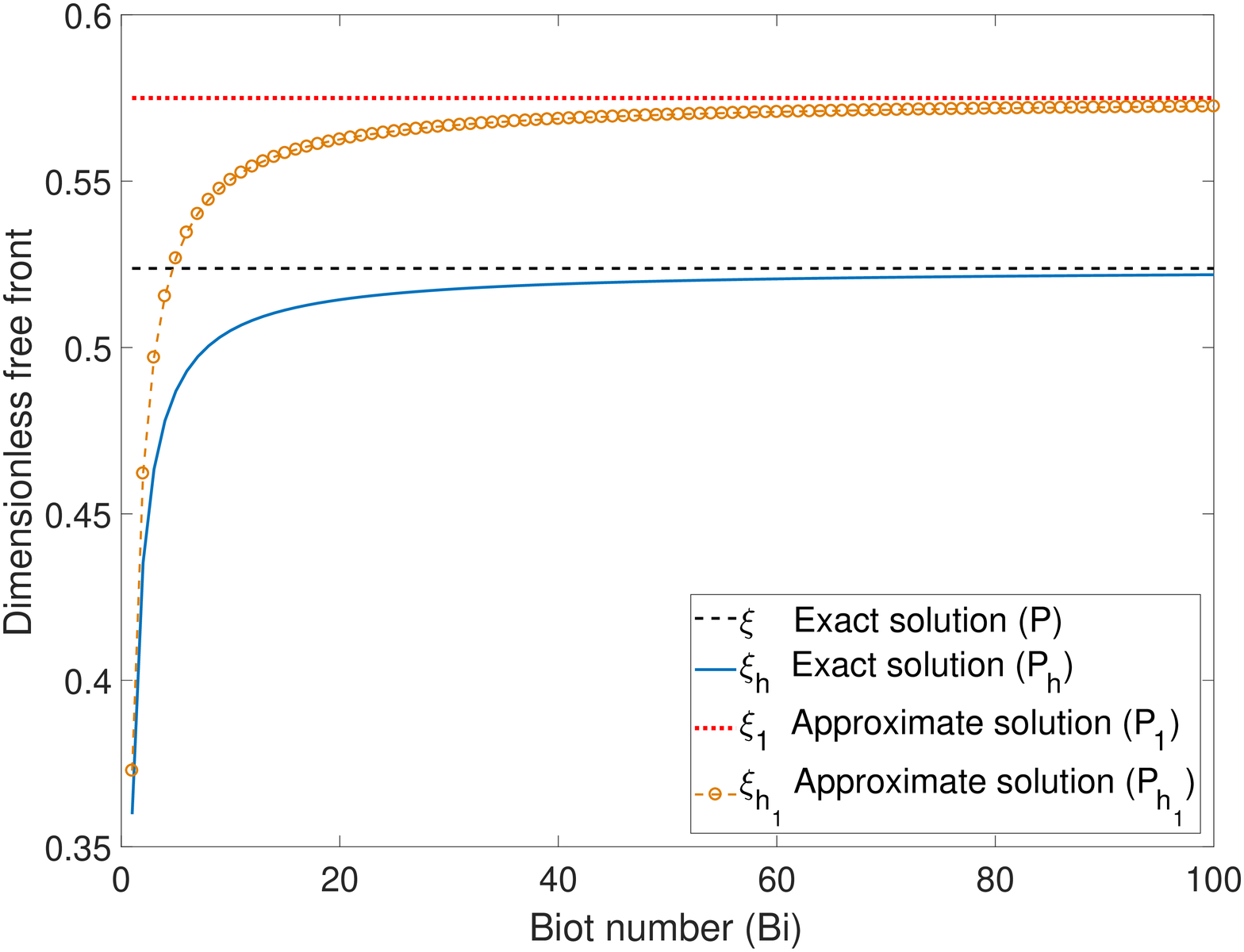} \\
{\scriptsize Figure 4: Plot of the dimensionless coefficients $\xi_h$ and $\xi_{h_1}$  against Bi number, for $\Ste=0.5$  and $\lambda=0.7$.}

\end{center}

\subsection{Approximate solution using a modified method of the classical heat balance method}
An alternative method of the classical heat balance integral method in order to solve the problem (P$_h$) proposes the resolution of the approximate problem (P$_{h_2}$) defined as follows:

Find the temperature $u_{h_2}=u_{h_2}(x,t)$ at the solid region $0<x<s_{h_2}(t)$ and the location of the free boundary $x=s_{h_2}(t)$ such that satisfy the following conditions: (\ref{TempFrontera})-(\ref{CondStefan})-(\ref{FrontInicial})-(\ref{CondConvecth})-(\ref{EcCalorAproxx}).

A solution to this problem will be an approximate one of the problem (P$_h$). Proposing the following quadratic temperature
profile in space:
\begin{equation}\label{Temp-uh2}
u_{h_2}(x,t)=A_{h_2}u_\infty\left(1-\frac{x}{s_{h_2}(t)}\right)+B_{h_2}u_\infty\left(1-\frac{x}{s_{h_2}(t)}\right)^2 ,\qquad  0<x<s_{h_2}(t)~, \quad  t>0~,
\end{equation}
the free boundary takes the form:
\begin{equation}\label{Front-sh2}
s_{h_2}(t)=2a\xi_{h_2}\sqrt{t},\qquad  t>0~,
\end{equation}
where the constants $A_{h_2}$, $B_{h_2}$ and $\xi_{h_2}$ will be determined from the conditions (\ref{CondStefan}), \ref{CondConvecth}) and (\ref{EcCalorAproxx}). Because of (\ref{Temp-uh2}) and (\ref{Front-sh2}), the conditions (\ref{TempFrontera}) and (\ref{FrontInicial}) are immediately satisfied. From conditions (\ref{TempFrontera}) and (\ref{EcCalorAproxx}) we obtain: 
\begin{equation}\label{Ah2}
A_{h_2}=\dfrac{2}{\Ste}\xi_{h_2}^2~,
\end{equation}
\begin{equation}\label{Bh2}
B_{h_2}=\dfrac{-\frac{2}{\Ste}\xi_{h_2}^3-\frac{1}{\Ste~\Bi}\xi_{h_2}^2+\xi_{h_2}}{\xi_{h_2}+\frac{1}{\Bi}}~.
\end{equation}

Since $A_{h_2}$ and $B_{h_2}$ are defined from the parameters $\xi_{h_2}$, $\Ste$ and $\Bi$, condition (\ref{CondStefan}) will be used to find the value of $\xi_{h_2}$. In this way, it turns out that $\xi_{h_2}$ must be a positive solution of the fourth degree polynomial equation:
\begin{eqnarray}\label{Xih2} 
z^4+\left(6\lambda+\dfrac{2}{\Bi}\right)z^3+\left(6+\Ste\right)z^2-\left(6\lambda\Ste+\dfrac{3}{\Bi}\right)z-3\Ste=0~, \qquad z>0~. 
\end{eqnarray}

Existence and uniqueness of solution for equation (\ref{Xih2}) can be easily seen by Descartes' rule. Therefore, the following theorem can be stated:

\begin{teo}\rm The solution of problem (P$_{h_2}$), for a quadratic profile in space, is given by (\ref{Temp-uh2}) and (\ref{Front-sh2}), where the constants $A_{h_2}$ and $B_{h_2}$ are defined by (\ref{Ah2}) and (\ref{Bh2}), respectively and $\xi_{h_2}$ is the unique positive solution of the polynomial equation (\ref{Xih2}).

In addition, the solution to problem (P$_{h_2}$) converges to the solution to problem (P$_2$) when $Bi \rightarrow \infty$.
\end{teo}

Comparisons between the exact solution $\xi_h$ with the approximate one $\xi_{h_2}$ are shown in Figure 5. We plot  them against Bi for $\Ste=0.5$ and $\lambda=0.7$. In order that the convergence of $\xi_h\to  \xi$ and $\xi_{h_2}\to \xi_2$ when $\Bi\to\infty$, could be appreciated, we also plot $\xi$ and $\xi_2$.

\begin{center}
\includegraphics[scale=0.27]{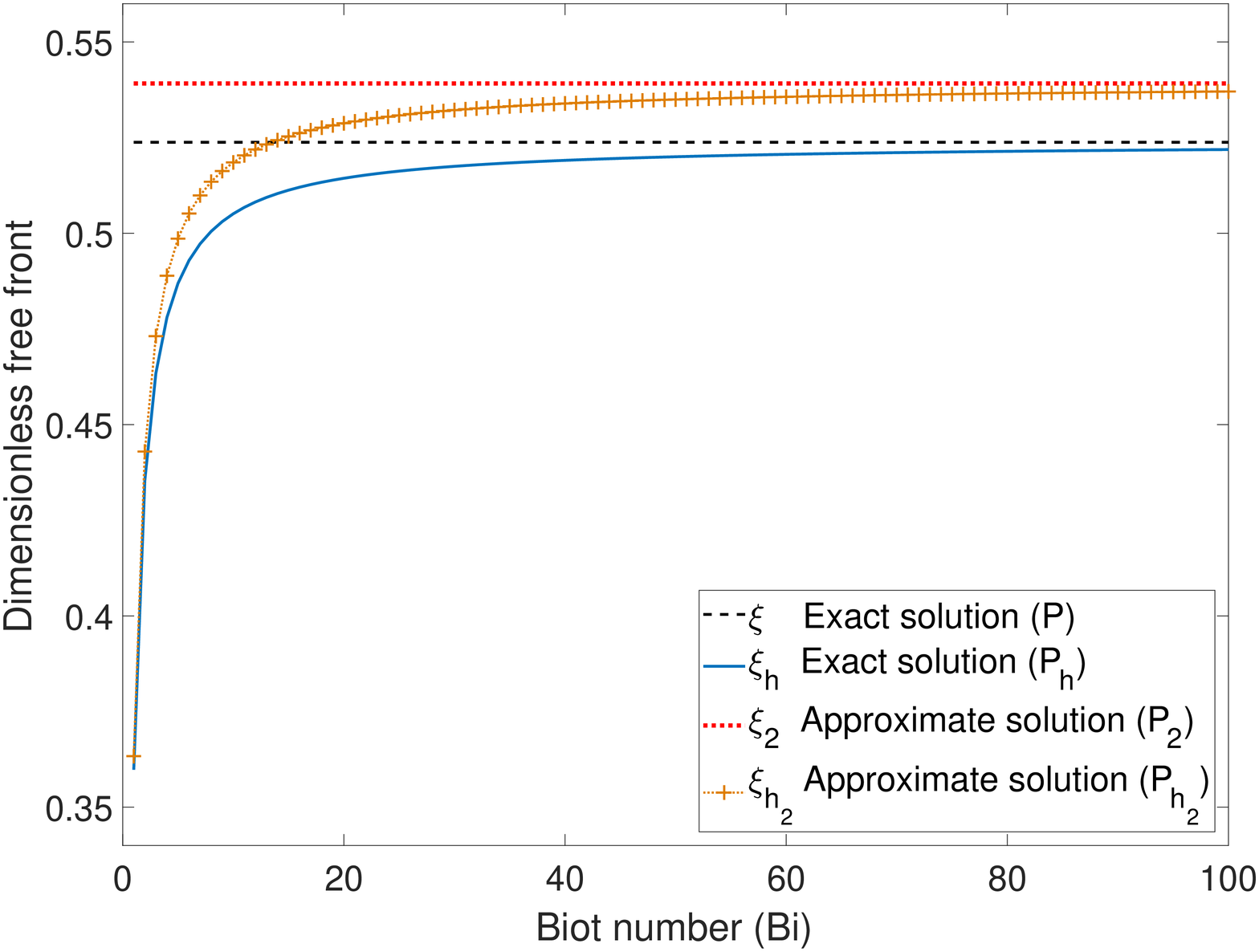} \\
{\scriptsize Figure 5: Plot of the dimensionless coefficients $\xi_h$ and $\xi_{h_2}$  against Bi number, for $\Ste=0.5$  and $\lambda=0.7$.}

\end{center}

\subsection{Approximate solution using the refined integral method}
The refined heat balance integral method in order to solve the problem (P$_h$), proposes de resolution of an approximate problem (P$_{h_3}$) formulated as follows:

Find the temperature $u_{h_3}=u_{h_3}(x,t)$ at the solid region $0<x<s_{h_3}(t)$ and the location of the free boundary $x=s_{h_3}t(t)$ such that satisfy the following conditions: (\ref{TempFrontera})-(\ref{CondStefan})-(\ref{FrontInicial})-(\ref{CondConvecth})-(\ref{CalorRIM}).

A solution to this problem will be an approximate one of the problem (P$_h$). Proposing the following quadratic temperature
profile in space:
\begin{equation}\label{Temp-uh3}
u_{h_3}(x,t)=A_{h_3}u_\infty\left(1-\frac{x}{s_{h_3}(t)}\right)+B_{h_3}u_\infty\left(1-\frac{x}{s_{h_3}(t)}\right)^2 ,\qquad  0<x<s_{h_3}(t)~, \quad  t>0~,
\end{equation}
the free boundary takes the form:
\begin{equation}\label{Front-sh3}
s_{h_3}(t)=2a\xi_{h_3}\sqrt{t},\qquad  t>0~,
\end{equation}
where the constants $A_{h_3}$, $B_{h_3}$ and $\xi_{h_3}$ will be determined from the conditions (\ref{CondStefan}), \ref{CondConvecth}) and (\ref{CalorRIM}). Because of (\ref{Temp-uh3}) and (\ref{Front-sh3}), the conditions (\ref{TempFrontera}) and (\ref{FrontInicial}) are immediately satisfied. From conditions (\ref{TempFrontera}) and (\ref{CalorRIM}) we obtain: 
\begin{equation}\label{Ah3}
A_{h_3}=\dfrac{2}{\Ste}\xi_{h_3}^2~,
\end{equation}
\begin{equation}\label{Bh3}
B_{h_3}=\dfrac{-\frac{2}{\Ste}\xi_{h_3}^3-\frac{1}{\Ste~\Bi}\xi_{h_3}^2+\xi_{h_3}}{\xi_{h_3}+\frac{1}{\Bi}}~.
\end{equation}

Since $A_{h_3}$ and $B_{h_3}$ are defined from the parameters $\xi_{h_3}$, $\Ste$ and $\Bi$, condition (\ref{CondStefan}) will be used to find the value of $\xi_{h_3}$. In this way, it turns out that $\xi_{h_3}$ must be a positive solution of the third degree polynomial equation:
\begin{eqnarray}\label{Xih3} 
-\left(6\lambda+\dfrac{1}{\Bi}\right) z^3-\left(6+\Ste\right)z^2+\left(6\lambda \Ste - \dfrac{3}{\Bi}\right)z+3\Ste=0~, \qquad z>0~. 
\end{eqnarray}
Clearly, by Descartes' rule of signs, we can assure that (\ref{Xih3}) has a unique positive solution. So, the following result holds:

\begin{teo}\rm The solution of problem (P$_{h_3}$), for a quadratic profile in space, is given by (\ref{Temp-uh3}) and (\ref{Front-sh3}), where the constants $A_{h_3}$ and $B_{h_3}$ are defined by (\ref{Ah3}) and (\ref{Bh3}), respectively and $\xi_{h_3}$ is the unique positive solution of the polynomial equation (\ref{Xih3}).

In addition, the solution to problem (P$_{h_3}$) converges to the solution to problem (P$_3$) when $\Bi \rightarrow \infty$.
\end{teo}

In Figure 6, the coefficient that characterizes the free boundary of the exact solution $\xi_h$ of problem (P$_h$) is compared with the coefficient $\xi_{h_3}$ that characterizes the free boundary of the  approximate problem (P$_{h_3}$), when we fix $\Ste=0.5$ and $\lambda=0.7$. We also show the value of $\xi $ and $\xi_{3}$ in order to visualize the mentioned convergence when $\Bi\to\infty$.

\begin{center}
\includegraphics[scale=0.27]{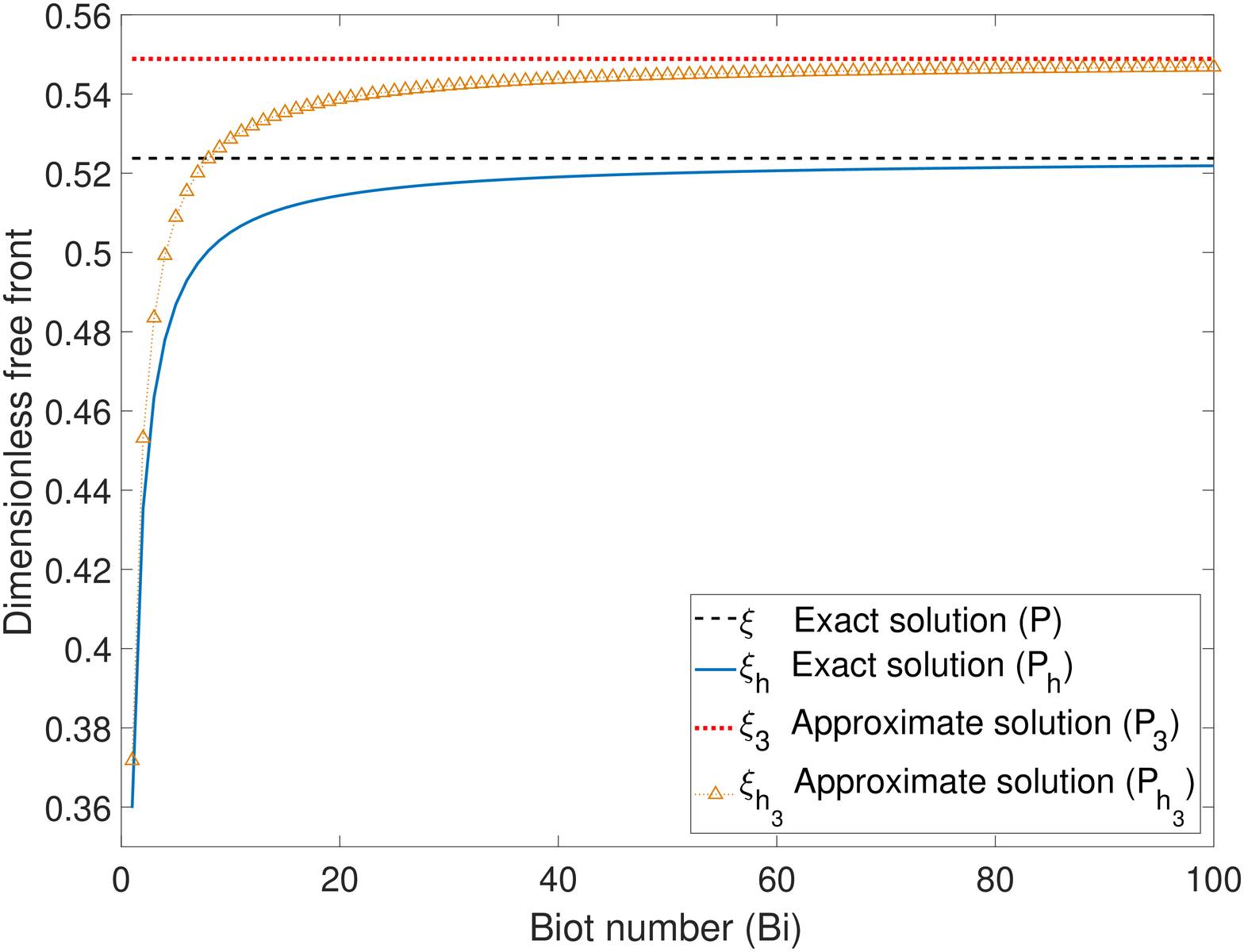} \\
{\scriptsize Figure 6: Plot of the dimensionless coefficients $\xi_h$ and $\xi_{h_3}$  against Bi number, for $\Ste=0.5$  and $\lambda=0.7$.}

\end{center}

\subsection{Comparisons between solutions}

Let us compare, for different Bi numbers, the numerical value of the coefficient $\xi_h$ given by (\ref{Xih})  and the approximations $\xi_{h_1}$, $\xi_{h_2}$ and $\xi_{h_3}$ given by (\ref{Xih1}), (\ref{Xih2}), (\ref{Xih3}), respectively.  In order to obtain which technique gives the best agreement, we display in Table 2, varying Bi between 1 and 100, the exact dimensionless free front, and its different approaches, showing also the percentage relative error committed in each case $E_{\text{rel}}(\xi_{h_i})=100 \frac{|\xi_{h_i}-\xi_{h}|}{|\xi_{h}|}$.

\begin{center}

\begin{footnotesize}

Table 2: Dimensionless free front coefficients and its relative errors.\\[0.15cm]
\begin{tabular}{cc|cc|cc|cc}
\hline
Bi     & $\xi_h$ (P$_h$)     &  $\xi_{h_1}$  (P$_{h_1}$) & $E_{\text{rel}}(\xi_{h_1})$       &   $\xi_{h_2}$  (P$_{h_2}$)   & $E_{\text{rel}}(\xi_{h_2})$      &    $\xi_{h_3}$  (P$_{h_3}$) & $E_{\text{rel}}(\xi_{h_3})$   \\
\hline

     1 &   0.3598  &  0.3729   \quad &  3.635 \%             & 0.3633  &  0.991 \%  &    0.3717  &  3.325 \% \\
   10  &   0.5051  &  0.5504   \quad &  8.965 \%             & 0.5185  &  2.657 \%  &    0.5286  &  4.655 \% \\
   20  &   0.5144  &  0.5626   \quad &  9.365 \%             & 0.5287  &  2.793 \%  &    0.5387  &  4.723 \% \\
   30  &   0.5175  &  0.5667   \quad &  9.501 \%             & 0.5322  &  2.840 \%  &    0.5421  &  4.745 \% \\
   40  &   0.5191  &  0.5687   \quad &  9.569 \%             & 0.5339  &  2.863 \%  &    0.5438  &  4.756 \% \\
   50  &   0.5200  &  0.5700   \quad &  9.610 \%             & 0.5350  &  2.878 \%  &    0.5448  &  4.763 \% \\
   60  &   0.5206  &  0.5708   \quad &  9.638 \%             & 0.5357  &  2.887 \%  &    0.5455  &  4.767 \% \\
   70  &   0.5211  &  0.5714   \quad &  9.657 \%             & 0.5362  &  2.894 \%  &    0.5459  &  4.770 \% \\
   80  &   0.5214  &  0.5719   \quad &  9.672 \%             & 0.5365  &  2.899 \%  &    0.5463  &  4.772 \% \\
   90  &   0.5217  &  0.5722   \quad &  9.684 \%             & 0.5368  &  2.903 \%  &    0.5466  &  4.774 \% \\
  100  &   0.5219  &  0.5725   \quad &  9.693 \%             & 0.5371  &  2.906 \%  &    0.5468  &  4.776 \% \\
    \hline
\end{tabular}

\end{footnotesize}

\end{center}

From Table 2, for the fixed values  $\Ste=0.5$ and $\lambda=0.7$,  we can appreciate that the error committed in each approximation increases when $\Bi$ becomes greater. We can notice that for the problems (P$_{h_1}$), (P$_{h_2}$) and (P$_{h_3}$) the  percentage errors do not exceed 9.693 \%,    2.906 \%  and \mbox{4.776 \%} respectively.

\section{Conclusion}

In this paper we have considered two different  Stefan problems for a semi-infinite material for the non classical heat equation with a source which depends on the heat flux at the fixed face $x=0$. We have defined the problem (P) with a prescribed constant temperature on $x=0$, which has been   studied in \cite{BrTa} and   where it was found a unique exact solution of similarity type. Also we have considered the problem (P$_h$)  with a convective boundary condition at the fixed face which was studied in this article, proving existence and uniqueness of an exact solution. The aim of this paper was to apply the classical heat balance integral method, an alternative of it and the refined integral method to those two problems in order to obtain approximate solutions. Comparisons with known exact solutions have been made in all cases and all solutions have been
presented in graphical form, providing an overview of popular approaches considered in recent literature. We emphasise that the fact of having the exact solution of problems (P) and (P$_h$) allows us to measure the percentage relative committed error by the approximate techniques applied throughout this paper. In all cases,  a dimensionless analysis was carried out by using the parameters: Stefan number (Ste) and the generalized Biot number (Bi), considering Stefan number up to 1 due to the fact that it covers most of the phase change materials.

We have obtained, for $\lambda=0.7$ that the best approximate solution to problem (P) was given by (P$_2$) obtaining a relative percentage error that does not exceed 5\%. Furthermore the best approximation to problem (P$_{h}$)  was obtained by (P$_{h_2}$)  obtaining a relative error of 2.9\%.
Therefore it can be said that in general the optimal approximate technique
for solving (P) and (P$_h$) was given by the alternative form of the heat balance integral
method, in which the Stefan condition is not removed and
remains equal to the exact problem.

In addition it was studied the case when Bi goes to infinity in the solution to the exact problem (P$_h$) an the approximate problems (P$_{h_1}$), (P$_{h_2}$), (P$_{h_3}$), recovering the solutions to the  exact problem (P) and the  approximate  problems (P$_1$), (P$_2$), (P$_3$). Some numerical simulations were also provided in order to visualize this asymptotic behaviour.

\section*{Acknowledgements}

The present work has been partially sponsored by the Project PIP No 0275 from CONICET-UA,
Rosario, Argentina, and ANPCyT PICTO Austral No 0090.

\small{

\end{document}